# Micro-Raman Spectroscopy of Mechanically Exfoliated Few-Quintuple Layers of $Bi_2Te_3$, $Bi_2Se_3$ and $Sb_2Te_3$ Materials

K. M. F. Shahil, M. Z. Hossain, V. Goyal and A. A. Balandin*

Department of Electrical Engineering

Materials Science and Engineering Program

Bourns College of Engineering

University of California - Riverside

Riverside, California 92521 U.S.A.

*Corresponding author (AAB): balandin@ee.ucr.edu ; http://ndl.ee.ucr.edu/

*Abstract*

Bismuth telluride ($Bi_2Te_3$) and related compounds have recently attracted strong interest owing to the discovery of the topological insulator properties in many members of this family of materials. The few-quintuple films of these materials are particularly interesting from the physics point of view. We report results of the micro-Raman spectroscopy study of the "graphene-like" exfoliated few-quintuple layers of $Bi_2Te_3$, $Bi_2Se_3$ and $Sb_2Te_3$. It is found that crystal symmetry breaking in few-quintuple films results in appearance of $A_{1u}$-symmetry Raman peaks, which are not active in the bulk crystals. The scattering spectra measured under the 633-nm wavelength excitation reveals a number of resonant features, which could be used for analysis of the electronic and phonon processes in these materials. In order to elucidate the influence of substrates on the few-quintuple-thick topological insulators we examined the Raman spectra of these films placed on mica, sapphire and hafnium-oxide substrates. The obtained results help to understand the physical mechanisms of Raman scattering in the few-quintuple-thick films and can be used for nanometrology of topological insulator films on various substrates.

## I. Introduction

Topological insulators (TI) constitute a new class of materials, which recently attracted significant attention of the physics and device research communities. The TI materials were shown to have an unusual insulating gap and exhibit quantum-Hall-like behavior in the absence of a magnetic field [1-8]. The strong spin-orbit coupling dictates robust surface states, which are topologically protected against back scattering from time-reversal invariant defects and impurities. It was suggested that TIs can be used for fault-tolerant quantum computing [7-8]. The unique properties of TIs show promise for the dissipation-less quantum electronics and spintronic devices operating at room temperature (RT). The thin films made of TI materials were proposed for applications in magnetic memory where write and read operations can be achieved by purely electric means [8].

It has been experimentally demonstrated that bismuth telluride ($Bi_2Te_3$) family of materials, e.g. $Bi_2Te_3$, $Bi_2Se_3$ and $Sb_2Te_3$, are TIs with energy gaps and gapless surface states consisting of a Dirac cone [4-6]. In addition to their new role as TIs, $Bi_2Te_3$ and related materials are among the best and widely used thermoelectric materials for near RT applications [9-13]. Apart from being thermoelectric materials, $Bi_2Se_3$ was also used in infrared detectors [14] while $Sb_2Te_3$ utilized in chalcogenide alloys as the phase changing material for information storage [15].

Despite increasing interest to $Bi_2Te_3$, $Bi_2Se_3$ and $Sb_2Te_3$ as TIs and their significance for thermoelectric industry, very few Raman spectroscopy studies of these materials were reported to date [16-22]. Most of the published Raman studies of these materials investigate bulk crystals or relatively thick film (~μm range), which are either amorphous or polycrystalline. The Raman spectroscopy data can provide useful structural information, crystalline phase, composition, and stoichiometry of the samples. It helps in understanding phonon dynamics and optical properties as well as serves as the means for determining the strain and compression in thin films and nanostructures. Most recently micro-Raman spectroscopy became valuable as a nanometrology tool for graphene. It allows one to count accurately and non-destructively the number of atomic planes in graphene samples [23], determine its quality and identify graphene on variable substrates and at different temperatures [24]. Raman spectroscopy was also used for the very first measurement of thermal conductivity of graphene [25].

Here, we report a systematic micro-Raman spectroscopy study of the thin crystalline films of $Bi_2Te_3$, $Bi_2Se_3$ and $Sb_2Te_3$. The samples for this study were prepared via the "graphene-like" mechanical exfoliation. We have recently successfully applied this technique for fabrication of the atomically-thin single-crystal films and ribbons of $Bi_2Te_3$ [26] and investigation of their electrical and electronic noise properties [27]. The developed exfoliation technique can be readily extended to other thermoelectric material systems [28]. The single-crystal $Bi_2Te_3$, $Bi_2Se_3$ and $Sb_2Te_3$ films with the thicknesses of just a few nanometers were prepared via the "graphene-like" exfoliation process. The resulting few-quintuple layer (FQL) films are well suited for both investigation of their physical properties and potential device applications.

In the present research we focused on FQL structures. Owing to the low thermal conductivity of $Bi_2Te_3$ and related materials a systematic Raman study of individual quintuples is complicated by a strong local heating and melting even at low laser excitation power. It has also been suggested that FQL are more promising as TI structures. While the band gap in a single quintuple is larger than in FQL, the latter has less coupling between the surface states of the top and bottom interfaces. The FQL $Bi_2Te_3$, $Bi_2Se_3$ and $Sb_2Te_3$ films are also more practical for thermoelectric applications. Taking the graphene [29] analogy even further we propose micro-Raman spectroscopy as a nanometrology tool for identification of FQL films of topological insulators and for assessing their quality [30]. Extending our previous investigation, in this paper we present a systematic study of the thin films made of several materials - $Bi_2Te_3$, $Bi_2Se_3$ and $Sb_2Te_3$ – which constitute some of the most promising TI and thermoelectric systems.

## II.  CRYSTAL STRUCTURE AND LATTICE VIBRATIONS

The $B_V$-$A_{VI}$ compounds (i.e. $Bi_2Te_3$, $Bi_2Se_3$, and $Sb_2Te_3$) exhibit a layered, rhombohedral crystal structure of the space group $R\bar{3}m(D_{3d}^5)$ [16-19]. These materials are built of anisotropic layers in which five atomic planes are covalently bonded to form a *quintuple* layer as shown in Figure 1 (a). A *quintuple* consists of five mono-atomic planes of $A_{VI}^{(1)}$-$B_V$-$A_{VI}^{(2)}$-$B_V$-$A_{VI}^{(1)}$. Here $A_{VI}$ can be either Te or Se, and $B_V$, either Bi or Sb. The superscripts on the $A_{VI}$ atoms designate the different positions within the fivefold layer. The 5-layer stacks are centro-symmetrical with respect to $A_{VI}^{(2)}$ which plays a role of an inversion center [18]. The conventional unit cell spans over three quintuple layers and each quintuple layer has a thickness of H~1 nm (see Figure 1 (a)). The quintuple layers are weakly bound to each other by the van der Waals forces that allow one to disassemble $B_V$-$A_{VI}$ crystal into its building blocks. In some cases, the atomic five-folds can be broken further into sub-quintuple leading to $B_V$-$A_{VI}$ atomic bi-layers and $B_V$-$A_{VI}$-$B_V$ atomic tri-layers. While the band gap in a single quintuple is larger than in FQL, the latter has less coupling between the surface states of the top and bottom interfaces.

The normal modes of vibration propagating along the trigonal $C_H$ axis involve motions of the entire planes of atoms, either parallel or perpendicular to $C_H$, and thus depend directly upon the inter-planar forces. At the Brillouin zone centre, there are four distinct representations, namely $2\Gamma_1^+, 2\Gamma_3^+, 3\Gamma_2^-$ and $3\Gamma_3^-$ [19]. These phonon modes are exclusively either Raman or infrared (IR) active due to the inversion crystal symmetry [18]. The $\Gamma_1^+$ and $\Gamma_3^+$, modes are Raman active, while the $\Gamma_2^-$ and $\Gamma_3^-$ modes of non-zero frequency are infrared active. So bulk crystals reveal 15 lattice vibration modes (phonon polarization branches). Three of these branches are acoustic and 12 are optical phonons. According to the group theory classification, 12 optical branches have $2A_{1g}$, $2E_g$, $2A_{1u}$, and $2E_u$ symmetry [18]. The corresponding Raman tensors of the Raman active modes are as follows [18]:

$$E_g = \begin{pmatrix} 0 & -c & -d \\ -c & 0 & 0 \\ -d & 0 & b \end{pmatrix} \text{ or } \begin{pmatrix} c & 0 & 0 \\ 0 & -c & d \\ 0 & d & 0 \end{pmatrix} \qquad A_{1g} = \begin{pmatrix} a & 0 & 0 \\ 0 & a & 0 \\ 0 & 0 & b \end{pmatrix} \qquad (1)$$

Here the letters "E" and "A" indicate the in-plane and out-of-plane ($C_H$ axis) lattice vibrations, respectively. The subscript "g" denotes Raman active while "u" stands for IR-active modes. The optical modes belonging to $A_{1u}$ and $E_u$ are allowed to be infrared active.

According to the selection rule given in Eq. (1), the off-diagonal Raman tensor components of the $E_g$ mode distinguish itself from $A_{1g}$ mode. This is because a phonon of this type possesses displacements in both x- and y- direction. In the $E_g^1$ and $A_{1g}^1$, modes the outer $B_V$-$A_{VI}^{(1)}$ pairs move in phase. Thus the $B_V$-$A_{VI}^{(2)}$ bonding forces will be primarily involved in these vibrations, whereas, in the $E_g^2$ and $A_{1g}^2$ modes the outer $B_V$ and $A_{VI}^{(1)}$ atoms move in the opposite phase, and are mainly affected by the forces between $B_V$ and $A_{VI}^{(1)}$ atoms. The nearest-neighbor distances between $B_V$ and $A_{VI}^{(1)}$ atoms are smaller than those between $B_V$ and $A_{VI}^{(2)}$ atoms [18]. For this reason, the $A_{1g}^1$ and $E_g^1$ modes occur at the lower frequencies than the $A_{1g}^2$ and $E_g^2$ modes.

## III. EXPERIMENTAL DETAILS

### A. Sample Preparation

The FQL samples of $Bi_2Te_3$ and related materials ($Bi_2Se_3$, $Sb_2Te_3$) were separated from the bulk samples using a micro-mechanical cleavage process similar to that used for exfoliation of single layer graphene [29]. The "graphene-like" mechanical exfoliation from bulk crystals allowed us to obtain the high-quality crystalline films, which was important for this study. The exfoliated films were examined using a combination of the optical microscopy, atomic force microscopy (AFM) and scanning electron microscopy (SEM) to obtain thickness of a few atomic planes. All FQL layers were exfoliated on $Si/SiO_2$ substrate. An optical image of the exfoliated $Bi_2Te_3$ FQL is shown in Figure 1 (b). The detailed sample preparation and identification techniques were described by some of us elsewhere [26-27, 30].

All exfoliated FQL films were investigated with the high-resolution field emission scanning electron microscope (SEM, XL-30 FEG) operated at 10-15 kV. Representative high-resolution SEM micrographs in Figure 1 (c-f) show SEM images of the reference bulk $Bi_2Te_3$ samples and FQL films of $Sb_2Te_3$, $Bi_2Te_3$, $Bi_2Se_3$ and respectively. All FGL films have the lateral sizes ranging from a few microns to tens of microns. Some FQL samples show a uniform surface and have correct geometrical shapes indicative of the facets and suggesting the high degree of crystalinity. The selected area electron diffraction of the crystalline structures of the layers was studied using FEI-PHILIPS CM300 transmission electron microscopy (TEM). The sample preparation for TEM inspection was carried out through ultrasonic separation of dissolved samples in isopropyl alcohol (IPA). The samples were then transferred onto carbon coated copper grids and studied under electron beam energy of 300 kV. The inset to Figure 1(g) shows a representative electron diffraction pattern of $Bi_2Se_3$ which indicates Bi-Se sample's perfect crystalline nature.

It is known that $Bi_2Te_3$ can be made either *n* or *p* type by changing the Bi/Te ratio. Even though the material that we use is pure $Bi_2Te_3$, the Bi/Te composition can deviate from the stochiometric due to Bi/Te differential surface tension effects and constitutional supercooling effects. The composition control is a serious technical problem because the thermoelectric properties of $Bi_2Te_3$ strongly depend on the composition. The defect chemistry in $Bi_2Se_3$ is dominated by the charged selenium vacancies, which act as the electron donors resulting in *n*-type behavior. In this work, the elemental composition of FQL Bi-Se films was studied by the energy dispersive spectrometry (EDS) using Philips XL-30 FEG field-emission system. The molar contents of Bi and Se were found to be ~38.98 % and 54.55 %, respectively. Figure 1 (g) presents EDS spectrum of $Bi_2Se_3$ FQL. A pronounced peak of Si indicates the electron beam penetration through thin FQL $Bi_2Se_3$ films. The diffraction patterns (Figure 1 (g) inset) of the crystalline structures of the layers were studied using the transmission electron microscopy (TEM).

The AFM studies were performed using a VEECO instrument with the vertical resolution down to ~0.1 nm in order to estimate the thickness of FQLs. The thickness of the atomic quintuple is H~1 nm which can be clearly distinguished with AFM by the step like changes in the cleaved layers. The high-resolution AFM images of the exfoliated $Bi_2Se_3$ FQL are shown in Figure 1 (h-i). The thickness of the films was measured along the line scan (see insets to the figures). The thickness profile is re-plotted (Figure 1 (h)) showing an 8-nm (~8 quintuple) step with respect to the substrate. In Figure 1 (i) another AFM image of $Bi_2Se_3$ is presented with the height profile and more uniform surface.

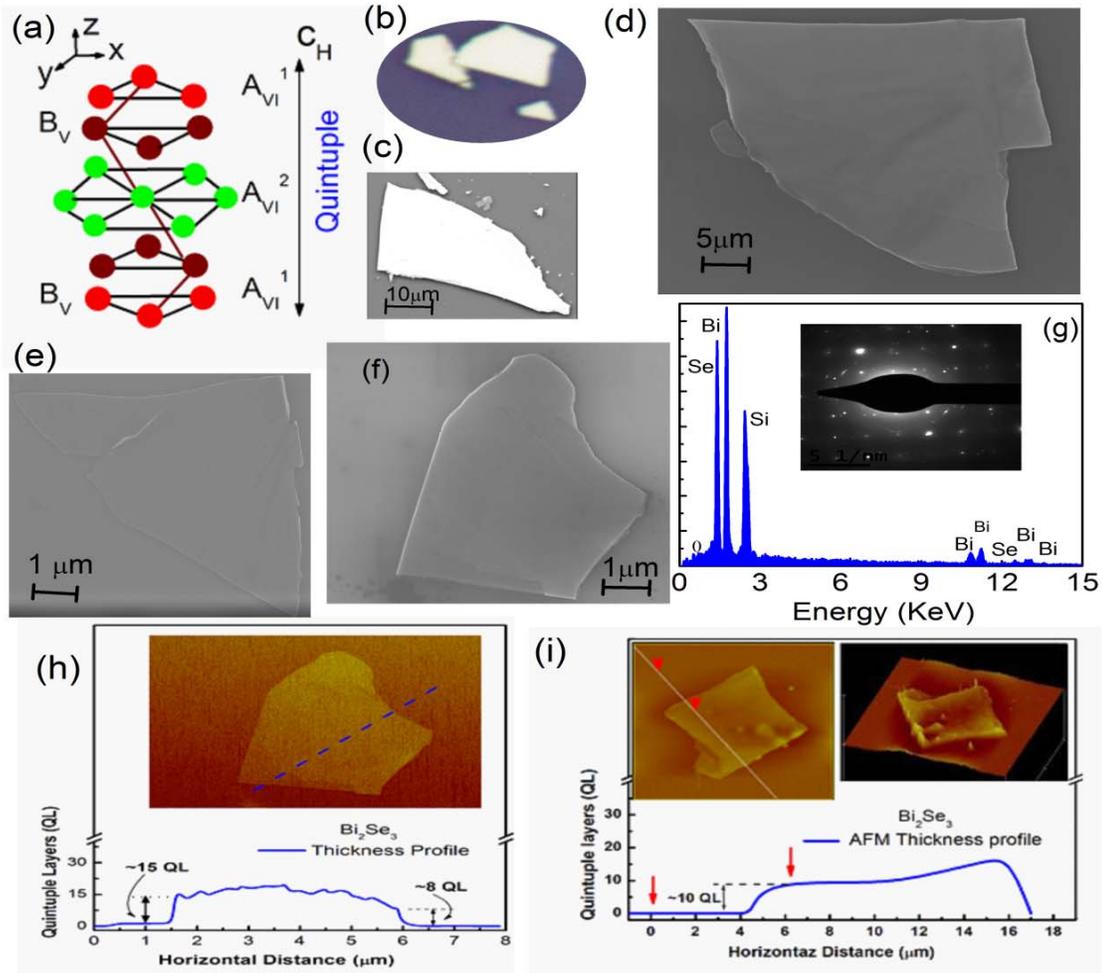

**Figure 1:** (Color online) (a) Crystal structure of the $B_V$-$A_{VI}$ compounds and a quintuple layer with the $A_{VI}^2$ ($Te^2$, $Se^2$) atoms as the inversion centers. SEM images of the bulk $Bi_2Te_3$ (c) and exfoliated $B_V$-$A_{VI}$ thin films with the large lateral dimensions showing (d) $Sb_2Te_3$, (e) $Bi_2Se_3$, and (f) $Bi_2Te_3$ flakes. (g) Structural and compositional characterization data showing EDS spectrum and electron diffraction pattern of $Bi_2Se_3$ FQL (inset) indicating the crystalline nature of the film. (h, i) AFM images of FQL of $Bi_2Se_3$.

### B. Raman Instrumentation

An upgraded Renishaw in Via micro-Raman spectrometer RM 2000 was used for this study. All spectra were excited at room temperature with laser light λ=488 nm and λ=633 nm recorded in the backscattering configuration through a 50× objective. With the 1800 lines/mm grating at 488 nm excitation and picking off the high resolution (-1 order) we can have the "hard-ware" spectral resolution of 1.35 cm$^{-1}$ (~1 cm$^{-1}$). It was software-enhanced to 0.5 cm$^{-1}$ so that we were able to see the difference between peaks (with plotted Lorentzians) down to approximately ~0.5 cm$^{-1}$. At the 633 nm excitation, the corresponding value of the spectral resolution was 0.63 cm$^{-1}$.

The Raman spectroscopic studies of thin films made from these materials are complicated due to the local heating effects since the materials have very low thermal conductivity and melting temperature [26]. The maximum excitation power of the Ar + laser with the wavelength of 488 nm used in this study was 10 mW. Approximately half of the excitation power reaches the sample surface after transmission through the optical system. The local laser-induced heating results in local melting or oxidation of FQL films [26, 30]. On the other hand, the measurement taken at insufficient excitation power produces low signal to noise (S/N) ratio since the Raman spectra of these low-bandgap materials generally show the lower Raman count owing to their metallic behavior. From the trial-and-error studies we established that the optimum excitation power in our setup was ~0.2 mW on the sample surface consistent. It provided a good signal-to-noise (S/N) ratio without damaging the FQL samples.

## IV. RESULTS AND DISCUSSION

### A. Non-resonant Raman Spectra

The measured non-resonant Raman spectra of Bi$_2$Te$_3$ FQLs are shown in Figure. 2. The spectra are recorded under 488-nm excitation. The observed four optical phonon peaks are identified (see Table 1). The frequencies of all these four zone-center regular Raman active modes are E$_g^1$ (TO) ~ 1.02 THz, A$^1_{1g}$ (LO) ~ 1.81 THz, E$_g^2$ (TO) ~ 3.02 THz and A$^2_{1g}$ (LO) ~ 4 THz [19]. These peaks are very close to the previously measured and assigned Raman peaks of Bi$_2$Te$_3$ bulk sample [18-21]. In addition to these peaks another peak, A$_{1u}$= 116.7 cm$^{-1}$ is found for the atomically thin FQL Bi$_2$Te$_3$ films. The A$_{1u}$ mode of the longitudinal optical (LO) phonons at 116.7 cm$^{-1}$ is IR active and corresponds to the zone-boundary phonon (Z point) of the frequency ~3.57 THz [18,19]. According to the theory [31], in crystals with the inversion symmetry, the IR-active modes like A$_{1u}$ must be odd parity while the Raman-active modes E$_g$, A$_{1g}$ must be even parity under inversion. The phonon displacement vector Q of an odd-parity phonon (IR) changes the sign under inversion, hence the Raman tensor ($d\chi/dQ$) of the odd-parity phonons in the centrosymmetric crystals must vanish [31] (see Eq. (1)). Thus, the odd-parity phonons (A$_{1u}$, E$_u$) do not show up in Raman spectra of bulk samples [20, 21] as long as the crystal retains its symmetry. We attribute the appearance of A$_{1u}$ mode in FQL to breaking of the crystal symmetry in the third dimension due to the limited thickness of FQL and presence of the interfaces.

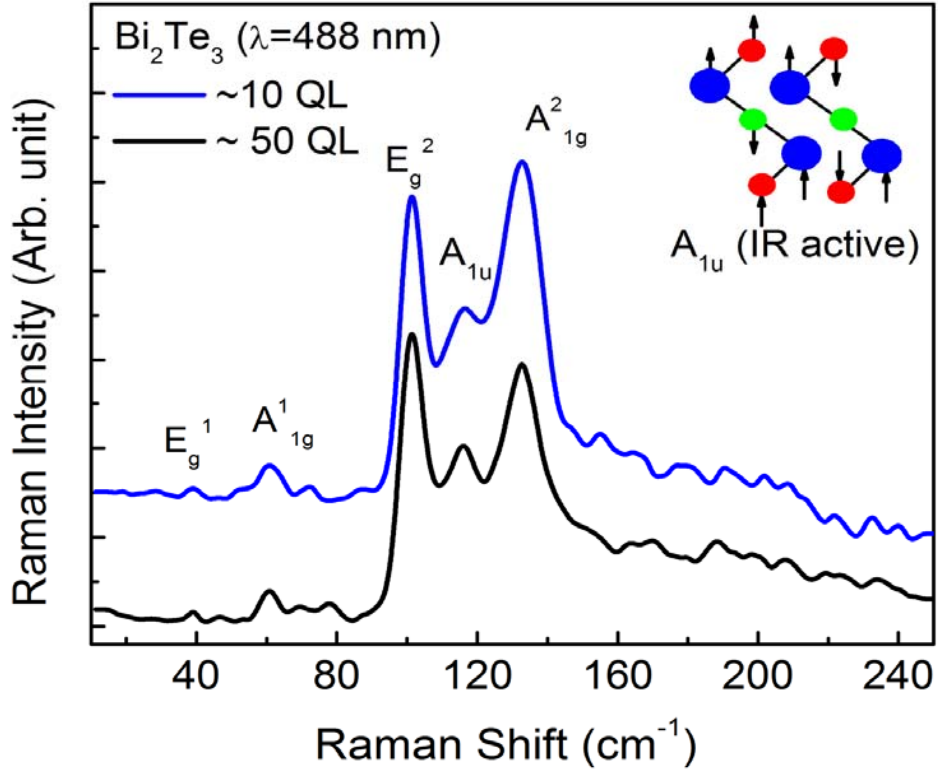

**Figure 2** (Color online): Raman spectrums of $Bi_2Te_3$ FQLs on $Si/SiO_2$ substrate. Spectra are taken from the spots with different thicknesses (~10 nm and ~50 nm) to demonstrate reproducibility. The displacement patterns of $A_{1u}$ phonon mode are shown in inset.

TABLE 1: Raman Peaks in FQL $Bi_2Te_3$ films

|       | $E^1_g$ | $A^1_{1g}$ | $E^2_g$ | $A_{1u}$ | $A^2_{1g}$ | Comments |
|-------|---------|------------|---------|----------|------------|----------|
| 10 nm | 38.5    | 60.7       | 101.4   | 116.5    | 132.9      | This work |
| 50 nm | 39      | 61.1       | 101.5   | 116      | 132.39     |           |
| Bulk  | 36.5    | 62.0       | 102.3   | -        | 134.0      | Ref. [17] |
| Bulk  | -       | 62.3       | 103.7   | -        | 134.2      | Ref. [21] |

We estimated the light penetration depth in our samples to be ~60–90 nm for 488 nm laser depending on the carrier concentration. One can calculate the penetration depth from the equation $\delta = 1/\sqrt{\pi f \sigma \mu}$, where $f$ is the frequency of light, $\sigma$ is the electrical conductivity of the materials and $\mu$ is the magnetic permeability. We assumed $\sigma$ in the range from $1.1 \times 10^5$ S/m to $5 \times 10^4$ S/m [26] and the calculated the penetration depth to be from ~60 nm to 90 nm at 488-nm laser light. This value correlates well with thickness $H$ when the 117 cm$^{-1}$ peak appears in FQL's spectrum. The onset of the $A_{1u}$ peak at the moment when the penetration depth becomes equal to that of the FQL thickness suggests that the loss of the translation symmetry at the FQL – substrate interface is the main mechanism of the crystal symmetry breaking.

Our results are in agreement with a recent computational study [32], which found that the crystal symmetry breaking in Bi$_2$Te$_3$ thin films should lead the Raman activity of $A_{1u}$. The *ab initio* calculation, which included the spin-orbit (SO) coupling, found that the single quintuple-layer films and the bulk materials have different symmetries [32]. The film has P-3m$^1$ symmetry while the bulk has R-3m symmetry. It was also noted that because of their limited thickness the films have no translational symmetry in the third dimension. The loss of the translational symmetry in films was considered to be one of the crystal breaking mechanisms. Another possible mechanism for very thin films identified in Ref. [32] was a strong inharmonic potential that exists around Bi atoms in the single quintuple films, which can be one of the symmetry breaking mechanisms caused by SO interactions. We also note that a single quintuple is inversely symmetric. Breaking of the individual quintuple layers to sub-quintuples, which in principle is possible, particularly on the film surface, could be an extra mechanism leading the crystal symmetry breaking and appearance of $A_{1u}$ peak. The effect from such sub-quintuple breaking will become stronger as the thickness of the films and, correspondingly, the Raman interaction volume, decreases.

As evidenced from ARPES measurements, an energy gap exists below a thickness of six quintuple layers of Bi$_2$Se$_3$ [33] and the films preserve gapless surface states when the thickness is above six QL. On the other hand, a significant improvement in thermoelectric applications could be achieved in 2D structure where electrons and holes are strongly confined in one or two dimensions. With this in mind, we calibrated the intensity ratio of highest-frequency $A_{1g}$ (i.e. $A^2_{1g}$ ~ 4 THz) mode to that of Eg$^2$ mode (Eg$^2$ the most pronounced feature in the spectrum) shown in Figure 3. We found that the intensity ratio grows as the thickness decreases from the bulk to a few quintuple layers. It is reasonable to assume that the out-of-plane vibrations will be less restrained in a four-quintuple film than in bulk, which may lead to larger amplitudes of vibrations. According to a theoretical calculation [32], there is another infra-active mode, $A_{2u}$ around ~3.97 Thz. This peak may also become Raman active due to the symmetry breaking. However, it is difficult to distinguish these two peaks experimentally because they are too close to each other. The strength of the Raman intensity of $A^2_{1g}$ mode (~4 THz) can increase due to the activation of another IR mode, which is close to the $A^2_{1g}$ mode frequency.

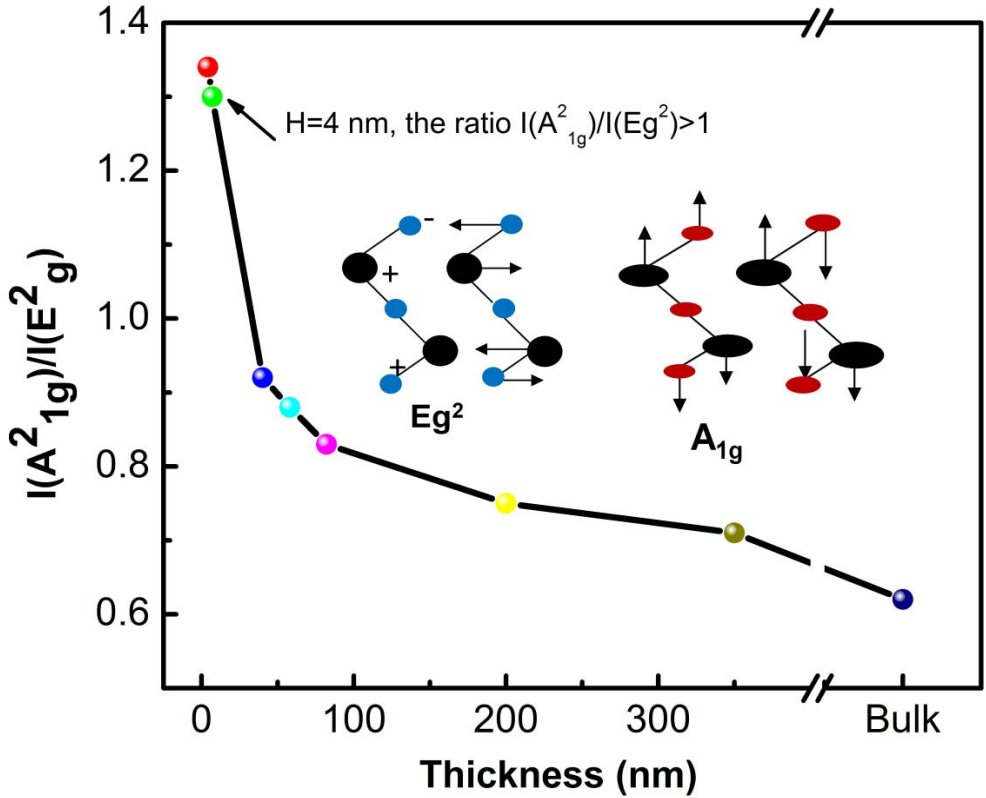

**Figure 3** (Color online): Calibrated Intensity ratio of Out of plane mode ($A^2_{1g}$) to that of in plane mode ($E^2_g$). The inset is the schematic of vibrational modes.

In this case, a combination of two Raman-active ($A^2_{1g} + A_{2u}$) modes, with an average frequency of ~4.03 THz will appear as the Raman signature around ~132 cm$^{-1}$ with increased intensity. The discussed modifications of the characteristic peaks in Raman spectra of FQL Bi$_2$Te$_3$ with the thickness create a basis for the use of micro-Raman spectroscopy as a nanometrology tool for characterization of these unique structures with the potential for practical applications.

We now turn to the analysis of Raman spectrum of FQL $Bi_2Se_3$ films. Figure 4 shows Raman spectra of $Bi_2Se_3$ FQLs taken in the geometry $z(xx)\bar{z}$ i.e. incident light shines along z-direction and both the incident and scattered light are polarized along x-direction. There are three characteristic peaks within the scanned frequency range: ~71 cm$^{-1}$, ~131 cm$^{-1}$ and ~173 cm$^{-1}$ which are identified in Raman spectra.

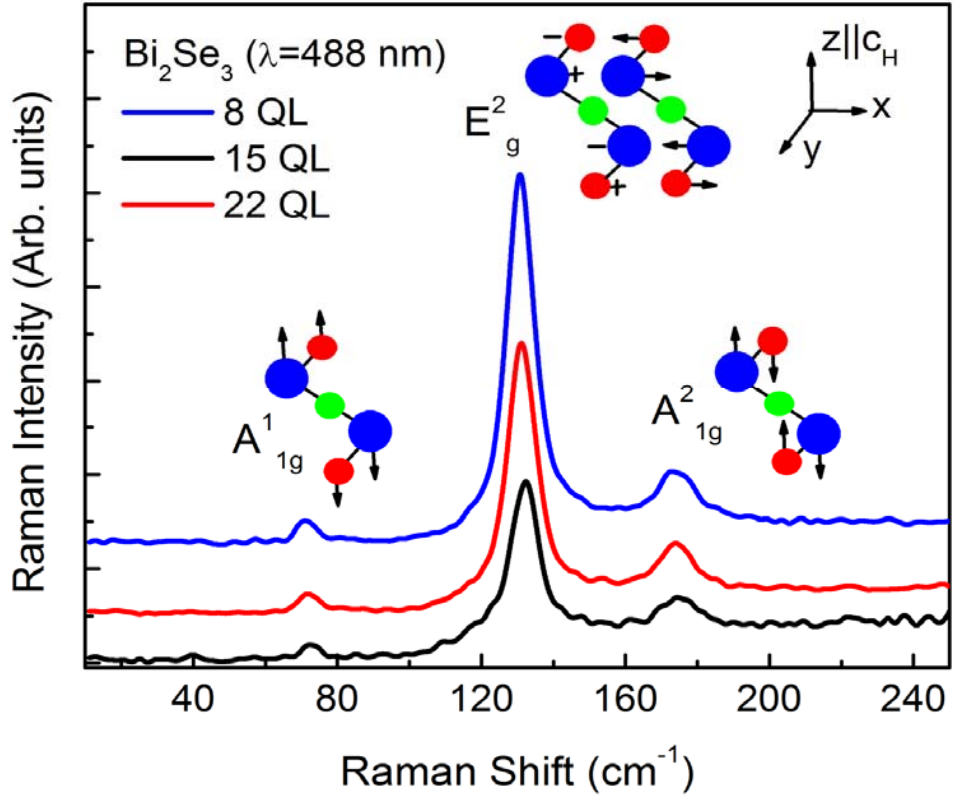

**Figure 4** (Color online): Raman spectra of $Bi_2Se_3$ FQLs in the 10-250 cm$^{-1}$ region. Three spectra are taken at different thickness of the sample. The displacement patterns of $E^2{}_g$, $A^1{}_{1g}$ and $A^2{}_{1g}$ phonon modes are shown in inset. The light is incident along z and polarized along x directions.

From the selection rules given in Equation 1 it follows that the peak at 131.5 cm$^{-1}$ has to be assigned to the $E^2{}_g$ mode whereas the other two (71 cm$^{-1}$ and 171.5cm$^{-1}$) corresponds to $A_{1g}$ mode. The observed frequencies are well comparable with the previously reported experimental and calculated phonon vibration modes of $Bi_2Se_3$ [18] (see Table 2). If polarization of x-direction is present ($z(xx)\bar{z}$ geometry), one will observe phonons denoted by $A^1{}_{1g}$, $A^1{}_{2g}$ and $E^2{}_g$. If one measures scattered light with a polarization in the y direction, only $E_g{}^2$ will appear in the Raman spectrum [18]. This can be understood by considering the nature of the atomic displacements associated with the phonon modes. If a phonon mode $A_{1g}$ is excited, the tensor implies that the polarization induced by the incident electric field $\xi_0 = (\xi_{x0}, \xi_{y0}, 0)$ has the same direction as $\xi_0$.

The $A_{1g}$ type phonon has $d\chi_{xy}/dQ = 0$. On the other hand, a phonon of the type $E_g$ possesses displacements in both x- and y- directions i.e., in plane vibration. An incident electric field $\xi_{x0}$ thus induces polarization changes in both directions and the scattered light contains polarization components in both these directions, i.e., $d\chi_{xx}/dQ \neq 0, d\chi_{xy}/dQ \neq 0$.

TABLE 2: Raman peaks in FQLs $Bi_2Se_3$ film

| Thickness | $A^1_{1g}$ | $E^2_g$ | $A^2_{1g}$ | Comment |
|---|---|---|---|---|
| ~8 nm | 70.6 | 131 | 173.4 | |
| ~15 nm | 72.4 | 132 | 174.7 | Our work |
| ~22 nm | 71.5 | 131.2 | 174.2 | |
| Bulk | 72 | 131.5 | 174.5 | Ref.18 |
| Calculated | 67.5 | 121 | 157.5 | |

For $Bi_2Se_3$ crystal, the Raman bands appear in the higher frequency range than that of $Bi_2Te_3$. This is due to the stronger bonding forces compared to $Bi_2Te_3$ and $Sb_2Te_3$. This fact is also supported by the small atomic distances in $Bi_2Se_3$ [18]. In addition, the Se atom is lighter than Te atom so that when laser light interacts with the material the Se atoms vibrate stronger producing Raman bands at a higher frequency. Another important observation is that there is no IR mode in the spectra of FQLs $Bi_2Se_3$ films as compared to that of $Bi_2Te_3$ films. According to Ref. [18], in $Bi_2Te_3$ crystals both $E \perp C$ and $E \parallel C$ polarizations are allowed to interact. However, in $Bi_2Se_3$, only $E \perp C$ polarization ($E_u$ mode) on the cleavage planes is possible. As a result, the $E_u$ mode, which is in-plane vibration, does not reveal itself even when the thickness of exfoliated $Bi_2Se_3$ goes down to a few quintuple layers along the $C_H$ axis.

Let us discuss now the non-resonant Raman spectrum of antimony telluride, which is shown in Figure 5. The calculated frequencies of the Raman active phonon modes for $Sb_2Te_3$ crystal were reported in Ref. [22]. For convenience, we listed them in Table 3. The calculations were preformed within the framework of density functional perturbation theory [22]. Comparing our results with the calculated values, we can conclude that in $Sb_2Te_3$ FQL the peak at ~35 cm$^{-1}$ corresponds to $E_g^1$ (TO), ~70 cm$^{-1}$ corresponds to $A_{1g}^1$ (LO), ~114 cm$^{-1}$ corresponds to $E_g^2$ (TO) and ~165 cm$^{-1}$ corresponds to $A_{1g}^2$ (LO). The agreement with the calculative data is good.

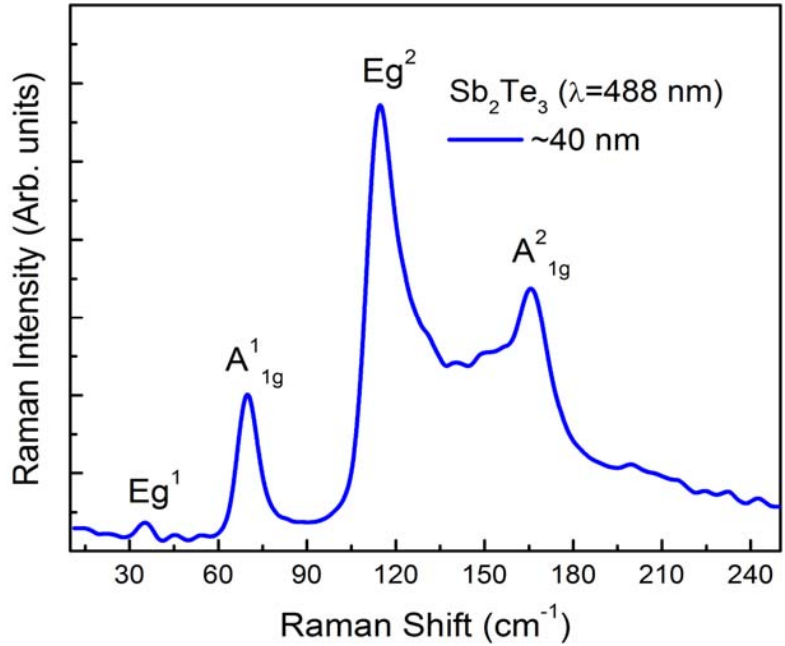

**Figure 5** (Color online): Raman spectra of sub-40 nm $Sb_2Te_3$ at 488-nm laser excitation.

TABLE 3: Raman peaks in FQLs $Sb_2Te_3$ film

|  | $E_g^1$ | $A_{1g}^1$ | $E_g^2$ | $A_{1g}^2$ | Comment |
|---|---|---|---|---|---|
| Sub 40 nm | 35.1 | 69.8 | 114.5 | 167.5 | Our work |
| Calculated | 46 | 62 | 113 | 166 | Ref. 22 |

### B. Resonant Raman Scattering

Raman spectroscopy can provide additional information if the resonant Raman scattering (RRS) occurs. When the excitation line is tuned into an electronic absorption band, some of the Raman bands, which are related to the electronic transition that is responsible for the absorption, will be greatly enhanced. This means that the equilibrium conformation of the molecule is distorted along the normal coordinate of the given Raman line in the transition from the ground to the excited electronic state. Theoretically, a change in the polarizibility $(\alpha_{p\sigma})_{mn}$ due to the electronic transition ($m \rightarrow e \rightarrow n$) can be written as [31]

$$(\alpha_{p\sigma})_{mn} = \frac{1}{h} \sum \left( \frac{M_{me} M_{en}}{\upsilon_{em} - \upsilon_0 + i\Gamma_e} + \frac{M_{me} M_{en}}{\upsilon_{en} + \upsilon_0 + i\Gamma_e} \right) \quad (2)$$

where $m$, $n$, $e$ are the electronic states, $\upsilon_{en}$ and $\upsilon_{em}$ are the frequencies corresponding to the energy differences between the states and incident laser beam frequency $\upsilon_0$. As $\upsilon_0$ approaches $\upsilon_{em}$, the denominator of the first term in the brackets of Eq. (2) becomes very small. Hence, this term – the resonance term – becomes so large that the intensity of the Raman band at $\upsilon_0 - \upsilon_{mn}$ increases strongly. This selectivity is important not only for identifying vibrations of this particular energy in a complex spectrum but also for locating its electronic transitions in an absorption spectrum. Therefore RRS has a wide range of application including determination of the electronic states.

The obtained Raman scattering spectra of FQL $Bi_2Te_3$ and $Bi_2Se_3$ taken at 633 nm wavelength are shown in Figure 6. Non-resonant Raman spectra taken at 488 nm is also shown for comparison. The same polarization setup (i.e. $z(xx)\bar{z}$) and the same conditions (i.e. accumulation time, laser power, and objective) were maintained for both 488 nm and 633 nm laser excitation. The strong resonance was observed as the excitation laser energy (1.96 eV) creates a vibrational state near the electronic state of the band structure.

This is exactly what is expected for both $Bi_2Se_3$ and $Bi_2Te_3$ since at $\Gamma$ point there exists a conduction band (electronic excited state) which lies at 633 nm (~1.96 eV) above the valence band maximum (ground state) [34].

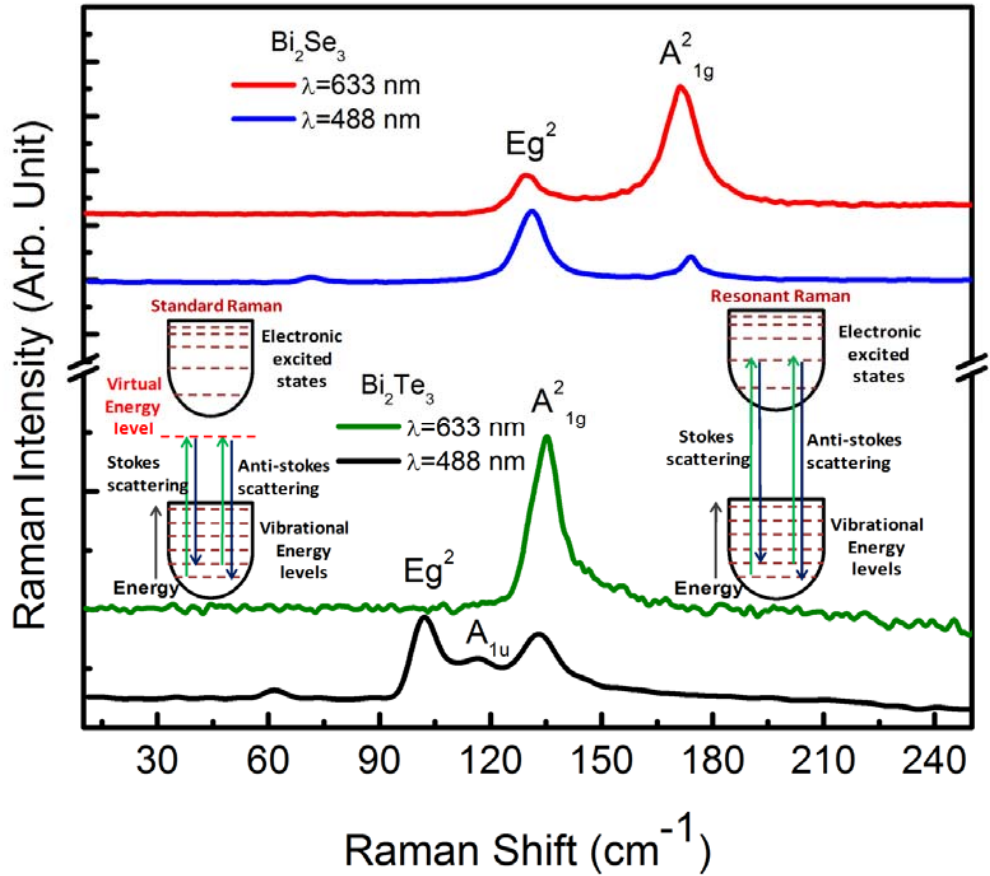

**Figure 6** (Color online): Resonant Raman spectra acquired at 633-nm (1.96 eV) excitation and non-resonant Raman spectra acquired at 488-nm (2.54 eV) excitation: (a) $Bi_2Te_3$ (b) $Bi_2Se_3$.

Another intriguing observation from Figure 6 is that at 488 nm (2.54 eV) the off-resonance excitations, the $E_g$ (TO) mode is the dominant feature in the spectra, while at the resonance with a laser excitation at 633 nm (1.96 eV) the $A_{1g}$ (LO) becomes the dominant mode. For $Bi_2Te_3$ a strong evolution of $A^2_{1g}$ mode (LO) is observed in Figure 6 and no $E_g^2$ mode (TO) can be observed under the resonant condition, i.e. 633-nm laser light. Similarly, in the case of $Bi_2Se_3$, the intensity of $A^2_{1g}$ mode becomes larger than that of $E_g^2$ mode. This evolution is analogous to the one observed by some of us in ZnO nanocrystals [35-36], where at the resonant conditions the LO is dominant and no TO phonons are observed. The dominance of LO modes can be explained by an analysis of how these phonons couple to the electronic systems. The LO phonon-electron interaction is mediated via Frohlich interaction [37] and the intensity of the LO modes might be strongly enhanced. On the other hand, the Frohlich coupling cancels for $E_g$ phonons. The properties of the longitudinal optical modes are of great interest because of the fact that their long range electrostatic field can couple and interact with electrons. This LO phonon-electron interaction can diminish the performance of optoelectronics devices [38] or, on the other hand, can be utilized in phonon engineering to create devices such as cascade lasers [39].

### C. Substrate Effects

In order to extend the use of Raman spectroscopy as the nanometrology tool for thin films, one needs to study how the Raman signatures of thin films are affected when they are placed on different substrates other than the regular $SiO_2/Si$. Previously we reported that the Raman phonon peaks undergo modification when graphene is placed on other substrates due to changes in the nature and density of the defects, surface charges and different strength of the graphene–substrate bonding [40]. After taking Raman spectra from FQL on the standard $SiO_2/Si$ substrate, we investigated $Bi_2Se_3$ placed on $HfO_2/Si$, sapphire, and Mica substrates (see Table 4). Figure 7 compares $Bi_2Se_3$ on different substrate ($HfO_2/Si$, MICA and sapphire) at 488-nm laser excitation. For sapphire at 488-nm light, the $A^2_{1g}$ mode shows a blue shift of 4 cm$^{-1}$.

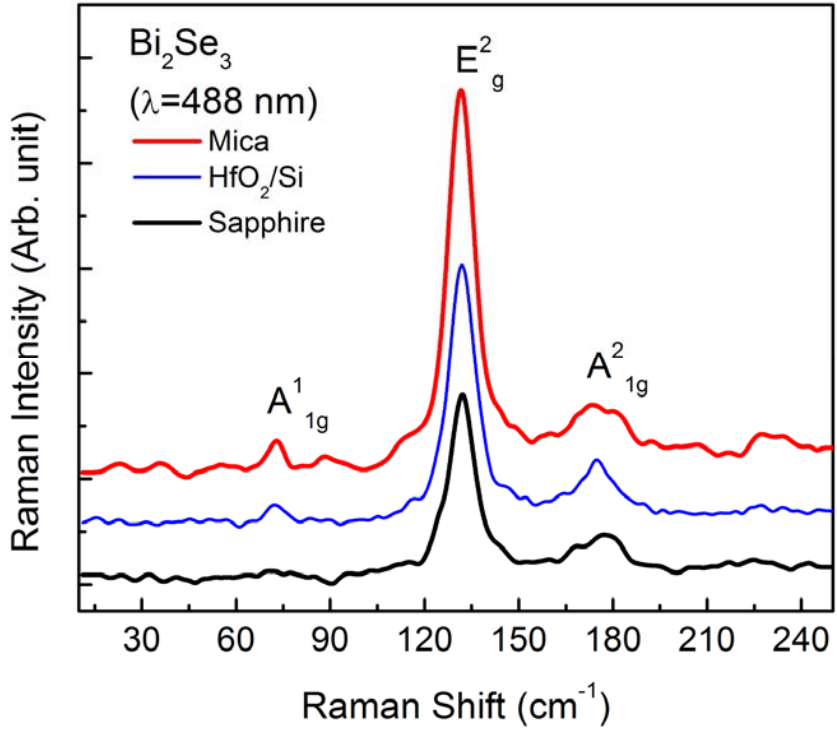

**Figure 7** (Color online): Raman spectra of $Bi_2Se_3$ FQLs on different substrates.

This could be due to the compressive strain. One can also see that at 488 nm the $A_{1g}^2$ mode splits. Previous Raman studies of our group of graphene layers on glass also indicated that in some instances G peak was split into doublets [40-41]. Random defects or charges on the film-substrate interface can explain the doublets. It is also worthwhile to mention that, the adherence of FQL to different substrates was similar. This conclusion could be important for further investigation of topological insulator on different substrates.

TABLE 4: Raman peaks in FQLs $Bi_2Se_3$ film on different substrate

| Substrate | $A_{1g}$ | $E_g$ | $A_{1g}$ |
|---|---|---|---|
| $SiO_2$/Si | 71.5 | 131.2 | 173.6 |
| Mica | 72.9 | 131.4 | 174.5 |
| $HfO_2$/Si | 71.9 | 131.7 | 175.0 |
| Sapphire | - | 133.3 | 177.1 |

## V. Conclusions

We presented results of the detail Raman studies of the few-quintuple-thick films of $Bi_2Te_3$, $Bi_2Se_3$ and $Sb_2Te_3$. The films were prepared by the "graphene-like" mechanical exfoliation from the bulk crystals. It was found that in some materials the crystal symmetry breaking in few-quintuple films results in appearance of $A_{1u}$-symmetry Raman peaks, which are not active in the bulk crystals. Our experimental observation is in agreement with the recent computational studies. The scattering spectra measured under the 633-nm wavelength excitation reveals a number of resonant features, which can also be used for analysis of the electronic and phonon processes in these materials. The obtained results help to understand the physical mechanisms of Raman scattering in the few-quintuple-thick films and can be used for nanometrology of topological insulator films on various substrates.

*Acknowledgements*

This work was supported, in part, the National Science Foundation (NSF) project on Coupled Charge and Spin Transport bin Topological Insulators and by the Semiconductor Research Corporation (SRC) – Defense Advanced Research Project Agency (DARPA) through FCRP Center on Functional Engineered Nano Architectonics (FENA). The authors acknowledge useful discussions with the former members of the Nano-Device Laboratory (NDL) – Dr. D. Teweldebrhan (Intel), and Dr. M Rahman (Intel) – for their help with EDS and TEM measurements